\documentclass[iop]{emulateapj}
\usepackage{natbib}

\def\kepler{{\slshape Kepler}}

\begin{document}
\title{The Kepler Dichotomy among the M dwarfs: \\Half of Systems Contain Five or More Coplanar Planets}

\author{Sarah~Ballard\altaffilmark{1,2} \&
John~Asher~Johnson\altaffilmark{3}}

\altaffiltext{1}{University of Washington, Seattle, WA 98195, USA; sarahba@uw.edu}
\altaffiltext{2}{NASA Carl Sagan Fellow}
\altaffiltext{3}{Harvard-Smithsonian Center for Astrophysics, Cambridge, MA 02138}


\begin{abstract}
We present a statistical analysis of the \kepler\ M dwarf planet hosts, with a particular focus on the fractional number of systems hosting multiple transiting planets. We manufacture synthetic planetary systems within a range of planet multiplicity and mutual inclination for comparison to the \kepler\ yield. We recover the observed number of systems containing between 2 and 5 transiting planets if every M dwarf hosts $6.1 \pm 1.9$ planets with typical mutual inclinations of 2.0$^{+4.0}_{-2.0}$ degrees. This range includes the Solar System in its coplanarity and multiplicity. However, similar to studies of \kepler\ exoplanetary systems around more massive stars, we report that the number of singly--transiting planets found by \kepler\ is too high to be consistent with a single population of multi--planet systems: a finding that cannot be attributed to selection biases. To account for the excess singleton planetary systems we adopt a mixture model and find that 55$^{+23}_{-12}$\% of planetary systems are either single or contain multiple planets with large mutual inclinations. Thus, we find that the so-called ``\kepler\ dichotomy'' holds for planets orbiting M dwarfs as well as Sun--like stars. Additionally, we compare stellar properties of the hosts to single and multiple transiting planets. For the brightest subset of stars in our sample we find intriguing, yet marginally significant evidence that stars hosting multiply--transiting systems are rotating more quickly, are closer to the midplane of the Milky Way, and are comparatively metal poor. This preliminary finding warrants further investigation.

%If we invoke another scenario in an  fashion to produce excess single transit systems, it must occur a fraction  of the time. We find that a model in which planetary formation occurs in two modes is 21 times likelier than a model with only one mode of planet formation. We compare the stellar properties of the hosts to single and multiple transiting planets, and investigate whether stellar rotation period, rotation amplitude, height from galactic midplane, or metallicity are predictive of one architecture over another. For the sample as a whole, only rotation period is significantly different between the two populations: we report that the M dwarf hosts to multiple transiting planets are rotating more quickly. However, when we consider the brightest subset of stars with $K$ magnitude less than 13.0 (80\% of the total sample), we can discriminate with 95\% confidence between the multiple and single transit systems in rotation period, galactic height, and metallicity. The multiply-transiting systems are rotating more quickly, are closer to the midplane of the Milky Way, and are comparatively metal poor. We consider possible links between these properties and the final planetary architecture. 
\end{abstract}

\section{Introduction}
While NASA's \kepler\ Mission was launched to investigate the frequency of planets orbiting Sun-like stars \citep{Borucki10}, the mission is foundational to our understanding of planet occurrence around the smallest stars \citep{Johnson14pt}. Though M dwarfs comprise less than 4\% of the \kepler\ targets (5500 stars of \kepler's 160,000 total, per \citealt{Swift13}), the \kepler\ planet yield encodes an occurrence rate of at least 1-2 planets per M dwarf with periods less than 150 days \citep{Dressing13, Swift13, Morton14b}. Our understanding of the mutual inclinations of exoplanets is also based, in large part, on results from the \kepler\ mission. Remarkably, 20\% of the planet host stars reported by \cite{Batalha13} host at least one transiting planet \citep{Fabrycky14}, including the first exoplanetary system discovered to have more than one transiting planet \citep[Kepler-9,][]{Holman10}.

%(Kepler-9,\citealt{Holman10}.

Despite the wealth of multi--planet systems discovered by other detection methods, the determination of the true mutual inclinations of the planets is limited to special cases: those orbiting pulsars \citep{Wolszczan08} and those with very strong dynamical constraints from radial velocity measurements \citep[e.g.][]{Laughlin05,Correia10}. For multi-planet systems discovered by transits, mutual inclinations can be measured on a system--by--system basis for those that transit starspots \citep{Sanchisojeda12} and those that transit one another \citep{Holman10b, Masuda13}. 

Other studies of multi--transiting systems have relied upon ensemble statistics to deduce the underlying mutual inclination distribution, including those of \cite{Lissauer11b}, \cite{Tremaine12}, \cite{Fang12}, and \cite{Fabrycky14}. All conclude that mutual inclinations less than 3$^{\circ}$ are consistent with the \kepler\ multi--planet yield, both in number of planets detected to transit and in the distribution of their transit durations. Indeed, flat and manifold architectures are necessary to recover the multi--planet statistics from \kepler. \cite{Lissauer11b} found that systems contain 3.25 planets per star (that is, 3 planets in 75\% of systems and 4 planets in 25\% of systems), with a mutual inclinations drawn from a Rayleigh distribution with $\sigma=2^{\circ}$, best reproduces the \kepler\ yield (excepting the small handful of systems with 5 and 6 transiting planets). \cite{Fang12}, similarly found that 75--80\% of planetary systems must host 1--2 planets, and 85\% percent of planetary orbits in multiple-planet systems are inclined by less than 3$^{\circ}$ with respect to one another. \cite{Swift13}, using the 5--planet system Kepler-32 as a template, thereby assuming 5 planets per star, found they could recover the multi-planet yield of \kepler\ systems orbiting M dwarfs with inclinations of 1.2$\pm0.2$ degrees. 

However, \cite{Lissauer11b} noted the puzzling finding that the best--fitting models to the \kepler\ yield underpredict the number of singly--transiting systems by a factor of two. \cite{Hansen13} reported a similar finding when comparing to their population synthesis model, which involved growing planetary architectures from multiple protoplanetary cores and then observing their final distributions. In both works, the authors posit a separate population of singly-transiting planets to explain the discrepancy. This feature of the \kepler\ multi--planet ensemble is coined ``the \kepler\ dichotomy.'' The mechanism responsible for producing an excess of singles is as-yet unclear. Either the primordial circumstellar disks of these stars produced less planets, or the resulting planets were scattered to larger mutual inclinations, ejected, or met their end in collisions with other planets or their host star. 

The conclusions of \cite{Morton14} favor the latter possibilities. They reported that that obliquity of planet host stars is larger for the singly-transiting planets, indicating a separate population of ``dynamically hotter'' systems. \cite{Johansen12} considered planetary instability as the responsible mechanism. They reported that planet-planet collisions in the typical packed \kepler\ architecture would occur with higher frequency than ejections or collisions with the host star. However, the planetary instability hypothesis is inconsistent with the resulting radius distributions of planets: there ought to be more small planets in the singly transiting systems, and more large planets in the multiply--transiting systems, for dynamical instability to be responsible. They go on to posit that the dichotomy must instead have arisen during the formation process itself, due to the effects of  massive planets on the protoplanetary disk. In this case, higher metallicity, which is responsible for increased giant planet occurrence  \citep[e.g.][]{Fischer05,Johnson09},  would be predictive of the final planetary architecture. 

If system architecture is, in fact, dependent on stellar metallicity, it is reasonable to investigate the effects of stellar mass as well, since both chemical composition and mass are two key properties of stars and planetary systems \citep[e.g.][]{Johnson10}. To this end, herein we investigate how many planets are required per star, and within what mutual inclination range, to specifically recover the \kepler's multi-planet yield around the least massive stars, the M dwarfs. We focus our study upon the smallest stars in the \kepler\ sample for several reasons. First, it is as-yet uncertain whether stellar mass plays a role in the \kepler\ dichotomy, which we can test with the sample of the smallest stars via comparison to previous studies of planets round Sun-like dwarfs. Secondly, the \kepler\ M dwarfs received a wealth of ground--based follow--up: nearly all of the KOIs orbiting M dwarf stars have published near--infrared or optical spectra, and often both, which facilitates accurate and precise estimates of stellar and hence planetary physical properties \citep{Muirhead12a, Ballard13, Mann13, Muirhead14}. The M dwarf sample is also less plagued by incompleteness: since the stars are so much smaller, the transits of smaller planets present a much larger signal (for example, transits that are a factor of 4 times deeper for a planet of a given size transiting M0V star compared to a Sun--like star), and some stars are near enough to have large proper motions that facilitate their validation \citep[e.g.][]{Muirhead12b}. 

M dwarf stars also present the best opportunities for future detailed  follow-up studies. We consider the case study of GJ\,1214\,b, a super--Earth discovered by the MEarth transit survey \citep{Charbonneau09}. Its orbital period of 1.6 days of and planet-to-star radius ratio of $\sim0.1$ are similar to those of hot Jupiters, but the small size and low luminosity of the host star renders GJ\,1214\,b both the smallest and coolest planet to receive detailed atmospheric study \citep{Bean10, Desert11, Berta12, Kreidberg14}.  NASA's all--sky Transiting Exoplaent Survey Satellite mission  will uncover a wealth of transiting planets after its launch in 2017 \citep{Ricker14}. However, the silhouette of an Earth-sized planetary atmosphere will be detectable only for the very brightest and smallest of these host stars, even with hundreds of hours of observation with the James Webb Space Telescope \citep{Deming09, Kaltenegger09}. Therefore, it is crucial to understand the architectures of planetary systems orbiting M dwarfs, partially in cases in which architecture bears upon potential habitability. These systems will very likely be the singular sites of atmospheric study of rocky planets.

In Section \ref{sec:data}, we describe our selection of the data set: the \kepler\ multiplicity yield of transiting planets orbiting M dwarfs. In Section \ref{sec:analysis}, we describe our procedure to generate synthetic planet samples to compare to those discovered by \kepler. We tune the parameters of our model, namely the number of planets per star and their mutual inclination distribution, to generate these synthetic planetary systems to fit for the number of planets per star. In Section \ref{sec:hosts} we investigate the stellar host properties, to test whether any parameter is predictive of planetary architecture. In Section \ref{sec:conclusions}, we summarize our conclusions and motivate future work. 

\section{Data Selection}
\label{sec:data}

We draw our sample of KOIs from the publicly available NASA Exoplanet Science Institute (NExScI) database\footnote{http://exoplanetarchive.ipac.caltech.edu/, accessed on 3 July 2013}.  We select our sample of M dwarf hosts from those characterized by \cite{Muirhead14}, which we supplement with those characterized by \cite{Mann14} that are hotter than 3950~K and cooler than 4200~K~\footnote{The Mann et al. sample contains stars hotter than 3950 K. Beyond this temperature threshold, the H$_{2}$O-K2 metric has saturated, per \citealt{Rojasayala12}, which is why they are not present in the Muirhead et al. sample}. From this list, we eliminate two eclipsing binaries identified in \cite{Mcquillan13a,Mcquillan13b} and \citet{Muirhead13}: KOIs 1459 and 256. We also remove KOIs 249 and 1725, which \cite{Muirhead14} determined to be blends, KOI 2626, which resides in a blended triple star system \citep{Star14}, and KOIs 531 and 3263, which \cite{Morton14b} found to have have false positive probabilities $>$0.95. In addition, one KOI present in \cite{Muirhead14}, KOI 1902.01, is now identified on the NExScI database to be a false positive, so we remove it. Our sample comprises 167 individual KOIs orbiting 106 stars: 71 host one KOI, 17 host two KOIs, 12 host three KOIs, 4 host 4 KOIs, and 2 host 5 KOIs (each of these has its own discovery paper: Kepler-32, described in \citealt{Swift13} and Kepler-186, described in \citealt{Quintana14}). 

Of the 109 stars included in our study, 84 are drawn from \cite{Muirhead14}, and 22 from \cite{Mann14}. \cite{Dressing13} also characterized the M dwarf sample of \kepler\ target stars with their broadband colors. In addition to target stars, there are 21 KOIs with properties reported in \cite{Dressing13} that are not present in \cite{Muirhead14} or \cite{Mann14}: of these, 9 are reported to be false positives in the NExScI database, and the remaining 12 have not received a planet candidate disposition. Finally, there are four KOIs that are dispositioned as candidates and likely cooler than 4200 K (from the broadband colors reported by \citealt{Dressing13}), but were not characterized in either manuscript: these are KOIs 2992, 3140, 3414, and 4087. All of the stars in the sample reside between the M4V--K7V spectral types\footnote{The KOI with the latest spectral type, KOI 4290.01, has an effective temperature of 3200 K, per \citealt{Muirhead14}}. 

\subsection{Investigation of Selection Biases}
\label{sec:bias}
\begin{figure}[h!]
\begin{center}
 \includegraphics[width=3in]{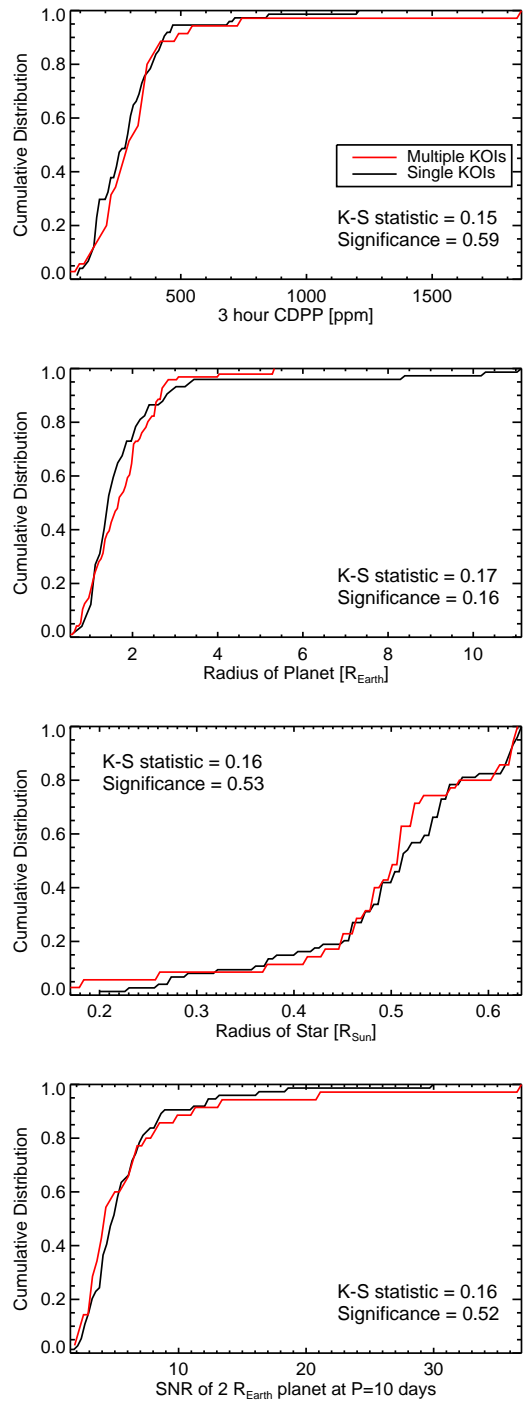} 
 \caption{Cumulative distribution in four parameters, from top to bottom: \kepler\ CDPP, planet radius, stellar radius, and a "typical" per-transit SNR. The latter corresponds to a 2 $R_{\oplus}$ planet in a 10-day orbit around each star, with CDPP noise scaled to the predicted transit duration at an impact parameter of zero. The red curve depicts the distribution for multiples, and the black for the singly-transiting KOI systems. The value of the K-S test and the significance are overplotted in each panel. There is a low probability in that the multiple systems are drawn from a different distribution than the singles in every parameter but radius, discussed in the text.}
  \label{fig:bias}
\end{center}
\end{figure}

We investigate the possibility that incompleteness or selection bias affects the observed multiplicity  of KOIs. We consider several explanations that are not astrophysical in nature, but rather produced by observational biases in the sample. First, the sample of \kepler\ M dwarfs is heavily weighted toward the largest and hottest of that spectral type, with most stars having spectral types M0--M1. Their larger size renders a given planet size less detectable because of the less favorable planet/star radius ratio. We have reason to be suspicious of our finding if single transit systems are simply likelier to reside around larger stars, since the smaller $R_{p}/R_{\star}$ means that more small planets {\it do} transit, but are eluding detection. Likewise, if noisier stars are overwhelmingly the likeliest hosts of single transit systems, we would need to consider the same possibility that noise is swamping transits that do exist, but don't lie above the detection threshold. In Figure \ref{fig:bias}, we depict the cumulative distribution for the single and multi-KOI systems for properties predictive of possible selection bias. These include the 3-hour Combined Differential Photometric Precision (CDPP), the stellar radius, and the planetary radius. Finally, we compare the per-transit signal-to-noise of a typical transit: that of a 2 $R_{\oplus}$ planet on a 10-day orbit. In a rough sense, this is translatable to the detectability of an average planet around each star. We overplot the Kolmogorov--Smirnov statistic for each parameter.

There exists no evidence that the singly-transiting KOI systems are drawn from a different distribution than that of the multiply-transiting KOI systems in CDPP, stellar radius, or average transit SNR. In these three cases, the test statistic is not in excess of the critical significance.
The most discrepant parameter between the two distributions is planetary radius. However, the discrepancy is distinct from the expectation for an observational bias, as follows. The three Saturn--sized and larger planet candidates ($>6$ $R_{\oplus}$) all orbit in singly-transiting systems: one of these systems is the only known Hot Jupiter orbiting an M dwarf: KOI 254 \citep{Johnson12}. The phenomenon of the singly-transiting systems hosting a population of larger planets is described in \citet{Johansen12} and \citet{Morton14}. But if we consider planets smaller than 6 $R_{\oplus}$ (that is, if we remove only these three planets), the planets in multiply--transiting systems are slightly larger than those in singly--transiting systems, which is the opposite of the predicted trend for a selection bias. This finding is apparent in the second panel of Figure \ref{fig:bias}. We report no evidence for selection bias producing an overly large sample of singly-transiting KOIs. 

Another important source of selection bias is the false--positive rate. We consider all KOIs in multiply-transiting systems to be authentic, following the reasoning described in detail in \cite{Lissauer12}, \cite{Rowe14}, and \cite{Lissauer14}. Therefore, false positives will appear as an excess of singly-transiting systems. \cite{Fressin13} reported an overall false--positive rate of 8.8$\pm$1.9\% for planets between 1.25--2 $R_{\oplus}$. Planets of this size comrpise 40\% of our sample. Another 30\% of the sample is smaller than 1.25 $R_{\oplus}$, which have a false positive rate of 12.3$\pm$3\%, and the remaining KOIs (nearly all between 2--4 $R_{\oplus}$) have a 6.7$\pm$1.1\% false positive rate \citep{Fressin13}. However, this sample of M dwarf KOIs is unusually pristine: we have already discarded 5 single KOIs as potential false positives because they are blends or eclipsing binaries, which is 6.3\% of the original sample of singles. \cite{Morton14b} individually calculated false positive probabilities (FPP) for 115 of the 167 KOIs in our sample. Of these, 87 have false positive properties $<0.05$, and 104 have false positive probabilities below 0.10. Of the 11 remaining KOIs with false positive probability greater than 0.1, 6 reside in multiply-transiting systems: in this work, we consider these to be authentic, as stated above. An additional KOI with FPP $>0.1$ is the {\it bona fide} hot Jupiter KOI 254.01 \citep{Johnson12}. In the event that the remaining 4 KOIs are all false positives, we adopt the conservative value of 5\% false positives among the single-KOI systems in hand. The findings we report in this work could only be falsified by false positive rates of 50\%, which we describe in the next section. 

\section{Analysis}
\label{sec:analysis}

\subsection{Methodology}
We compare the observed sample of KOIs to synthetic populations of exoplanets, which we manufacture across a grid of multiplicity and orbital mutual inclination, as we describe in detail in this section. We first assume a single type of planetary system architecture, in which each star hosts $N$ planets, with mutual inclinations drawn from a Rayleigh distribution with scatter $\sigma$. For the sake of comparison to the \kepler\ sample, we ask how many planets, $\mu$, of the total $N$ would be seen to transit, if the mean orbital plane of planetary systems is distributed isotropically. We define our model $\mu \equiv \mu_n(N,\sigma)$, which describes the expected number of transiting planets per star, given a parent population characterized by $N$ and $\sigma$. The index $n$ refers to each bin of stars having $\mu_n$ planets; $n$ does not refer to individual stars. In our analysis we consider indices $n$ running from 1--8.

We compare the model--predicted population for a given $\mu$ to the observed number $M_n$ of multiples with $n$ planets per star. Poisson counting statistics describe integer numbers of transiting planets, so we evaluate the likelihood of $N$ and $\sigma$ with a Poisson likelihood function, which is conditioned on the observed number of transiting planets in each bin, $M_n$, with the ensemble of bins given by $\{M\}$. Therefore, we can describe the likelihood $\mathcal{L} \equiv P(\{M\}|N,\sigma)$ of observing the distribution $\{M\}$, given some $N$ and some $\sigma$, by 

\begin{equation}
\mathcal{L} \propto \prod_{n}\frac{\mu(N,\sigma)^{M_n}e^{\mu(N,\sigma)}}{M_n!}.
\label{eq:bayes1}
\end{equation}

\noindent It remains for us to find the values of $N$ and $\sigma$ that maximize the likelihood of observing $\{M\}$.

\subsection{Modeling a single population of multi-planet systems}

We initially make the additional assumption of circular orbits for our simplest scenario. In order to evaluate how many planets we expect to transit, we use a Monte Carlo method to generate a synthetic transiting planet sample. We generate $10^5$ planetary systems for each value of $N$ and $\sigma$, allowing $N$ to vary from 1--8 planets in increments of one planet, and $\sigma$ to vary from 0$^{\circ}$--10$^{\circ}$ in increments of 0.1$^{\circ}$. In each scenario, we draw $N$ periods randomly from a flat distribution in $\log{P}$ space, ranging from 1--200 days. While other studies predict complex structure in the period probability distribution with several peaks (\citealt{Hansen13} from simulation work, for example), we assume the approximate flatness that \cite{Foreman14} reports from fitting to the \kepler\ data set. This probability does fall off slightly for periods $>50$ days for planets $<2$ $R_{\oplus}$, but is still consistent with flatness.

We then test each synthetic planetary system for Hill stability, using Equation 3 from \cite{Fabrycky14}. This expression defines the mutual Hill radius to be: 

\begin{equation}
R_{\rm H}=\left[\frac{M_{\rm in}+M_{\rm out}} {3M_{\star}}\right]^{1/3}\frac{(a_{\rm in}+a_{\rm out})}{2},
\end{equation}

\noindent where the stability criterion is satisfied if 

\begin{equation}
\label{eq:Hillineq}
\frac{a_{\rm in}+a_{\rm out}}{R_{\rm H}} > \Delta_{\rm crit},
\end{equation}

\noindent where $a_{\rm in}$ and $a_{\rm out}$ are the semimajor axes of the inner and outer planets, respectively, measured in AU. The critical separation is $\Delta_{\rm crit} = 2\sqrt{3}$ for adjacent planets. 
For systems with more than three planets, \cite{Fabrycky14} require that $\Delta_{\rm inner}+\Delta_{\rm outer} > 18$ for adjacent inner and outer pairs of planets.

If the set of planetary orbits violate these criteria, we reject that iteration and draw a new set of periods. We assign mutual inclinations from a Rayleigh distribution of angles with scatter $\sigma$. For each set of planets, we calculate the impact parameter $b$ of the planet in units of stellar radii, where the planet ``transits" if $b<1$. We record the number of planets that transit for each synthetic exoplanetary system. After 10$^{5}$ iterations, we produce the model histogram $\mu$ of the number of systems hosting $N$ planets, of which $n$ transit. 

\begin{figure}[h!]
\begin{center}
 \includegraphics[width=3in]{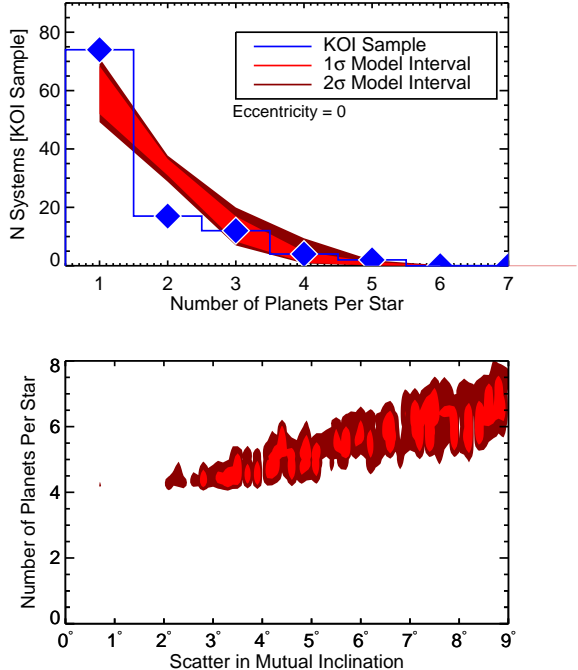} 
 \caption{{\it Top panel}: comparison of the \kepler\ multi-planet yield, in blue, to the best fit models at confidences of 1$\sigma$ and 2$\sigma$, depicted by the dark and light red regions respectively. {\it Bottom panel}: A contour map of the posterior distribution $P(\{ M\}|N,\sigma)$, with dark and light red corresponding again to 1$\sigma$ and 2$\sigma$ likelihoods.}
  \label{fig:eval_pdf_all}
\end{center}
\end{figure}

We then employ this empirical $\mu(N,\sigma)$ to evaluate Equation \ref{eq:bayes1} across a grid of $N$ and $\sigma$. We depict the resulting contour plot of $P(\{M\}|N,\sigma)$ in Figure \ref{fig:eval_pdf_all}. We also plot the \kepler\ yield and the range of best--fitting models, i.e. those that maximize the likelihood, in the 68\% and 95\% confidence intervals. A comparison of the best-fitting family of models to the \kepler\ sample shows that no set of $\{N,\sigma\}$ furnish a multi--transiting yield similar to \kepler's. The best--fitting models, which combine high multiplicity and a high degree of scatter in mutual inclinations significantly underpredict the number of singly transiting systems and overpredict the number of multiply transiting systems. The best--fitting models have $N=6\pm2$ planets and $\sigma=7.4^{+2.5}_{-4.6}$ degrees.

% JJ NOTE: I commented out this last part since it compares to a result of multis around FGK stars, which may be intrinsically different than those around M dwarfs.
%, which doesn't resemble a typical planetary system discovered by \kepler. \cite{Fabrycky14} found that the \kepler\ multiply-transiting systems were best fit with a mutual inclination of 1.0--2.2$^{\circ}$, well below the range of the best-fit model here.

\begin{figure}[h!]
\begin{center}
 \includegraphics[width=3in]{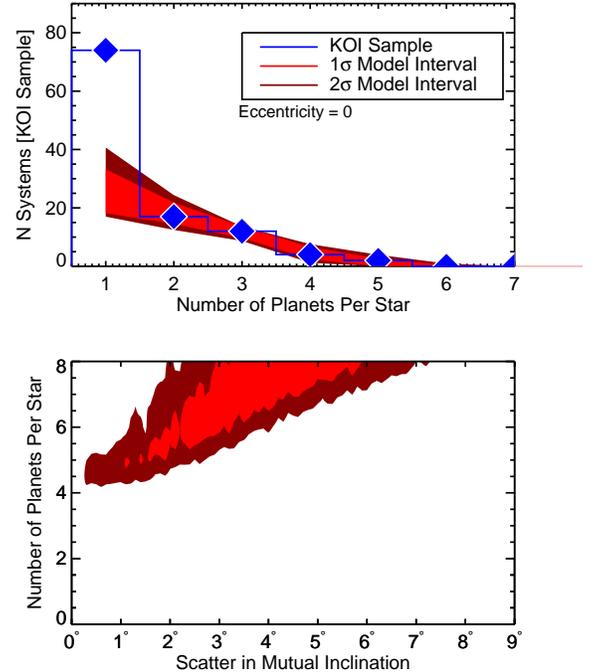} 
 \caption{{\it Top panel}: comparison of the \kepler\ multi-planet yield, in blue, to the best fit models at confidences of 1$\sigma$ and 2$\sigma$, depicted by the dark and light red regions respectively. This analysis compared only the yields for the multiple-planet systems, with the singly--transiting systems excluded from the likelihood calculation. {\it Bottom panel}: A contour map of the posterior distribution $P(\{ M\}|N,\sigma)$, with dark and light red corresponding again to 1$\sigma$ and 2$\sigma$ likelihoods.}
  \label{fig:eval_pdf_multionly}
\end{center}
\end{figure}

However, when we compare the predicted \kepler\ yields against only the set of multiply-transiting systems (i.e. excluding single--transiting systems, similarly to \citealt{Fabrycky14}) our results change significantly. First, we find good agreement between model predictions and the data, which is clear in Figure \ref{fig:eval_pdf_multionly}. Secondly, this posterior distribution is very dissimilar from the one derived from the full set of transiting planets (that is, including the singles). In order to reproduce \kepler's yield for systems with two or more KOIs, there must exist $>5$ planets per planet-hosting star, with a mutual inclination scatter of $\sigma=4.6^{+1.7}_{-3.0}$ degrees. This range is consistent, though less constraining, than the findings of \cite{Fabrycky14}. These parameter values also enclose the Solar System, with its $N=8$ planets and 2.1$^{+0.43}_{-0.21}$. The latter value is obtained by fitting a Rayleigh distribution to the mutual inclinations of the planets orbiting the Sun, as described in Section \ref{sec:hosts}. Of course, the Solar System planets span a much broader range of periods and masses than the typical system around an M dwarf. However, in this sense the M dwarf planetary systems can be viewed as scaled-down versions of the Solar System. 

We repeat the experiment with a modified and more physically--motivated assumption for eccentricity. \cite{Limbach14} employ the RV sample of exoplanets to study the eccentricity distribution of planets as a function of system multiplicity. They find that a mean eccentricity of 0.27 is representative for the sample of single-planet systems, but that this mean value decreases to 0.1 for systems with 5 or 6 planets (the mean eccentricity of the Solar System, as-yet the only known system with 8 planets, is 0.06). They provide the probability densities functions for eccentricity for multiplicities ranging from 1--8 planets per star, which they fit from the observed cumulative distributions in eccentricity of the known RV--detected exoplanets. 

We fold these density functions (Mary-Anne Limbach, 2014, private communication) into our machinery to simulate planetary systems, where the eccentricity of a planet in our simulated sample is drawn from the distribution corresponding to its number of neighboring planets. We assign the longitude of perihelion from a uniform distribution from 0$^{\circ}$ to 360$^{\circ}$ for each planet. We apply the Hill stability criterion using the same inequality in Equation \ref{eq:Hillineq}, but conservatively define $a_{\rm in}$ to be the apastron distance of the inner planet, and $a_{\rm out}$ to be the periastron distance of the outer planet. That is, we mandate that the closest possible approach of the two planets still satisfies the stability criterion. As in our analysis with circular orbits, we reject the synthetic system if the criterion is not satisfied for any one pair of adjacent planets, and draw a new set of periods. We consider mutual inclination as follows. We assume that equipartition holds, that is, if the Rayleigh distribution from which we draw inclination is defined by $\sigma=\sigma_{\rm inc}$ (with $\sigma$ expressed in radians), then the Rayleigh distribution from which we draw eccentricity should scale as $\sigma_{\rm e}=\eta\sigma_{\rm inc}$. 

Other investigations, such as \cite{Fabrycky14}, fit for $\eta$ from \kepler\ observables. From the set of observed transit durations, \cite{Fabrycky14} concluded that an $\eta=2$ fit the distribution best, but that values from $0>\eta>7$ were acceptable fits to the data. For this reason, we place no prior on $\eta$, but simply scale $\sigma_{\rm inc}$ with the $\sigma_{\rm e}$ from each distribution in eccentricity from \cite{Limbach14}. We assume that the assigned $\sigma$ value for each model applies for the flattest case (8 planets), and ought to be scaled up to produce reasonable models with fewer planets. For example, to produce $\mu(N=4, \sigma=4^{\circ})$ we consider the fact that systems with 4 planets are likeliest to have an average eccentricity 0.11, 2.3 times larger than systems with 8 planets (with likeliest eccentricity of 0.048). So, in a universe in which the flattest systems (N=8 planets) have a $\sigma$ of $4^{\circ}$, and those with $N=4$ planets should have a sigma 2.3 times larger, or $\sigma = 9.2^{\circ}$. 

\begin{figure}[h!]
\begin{center}
 \includegraphics[width=3in]{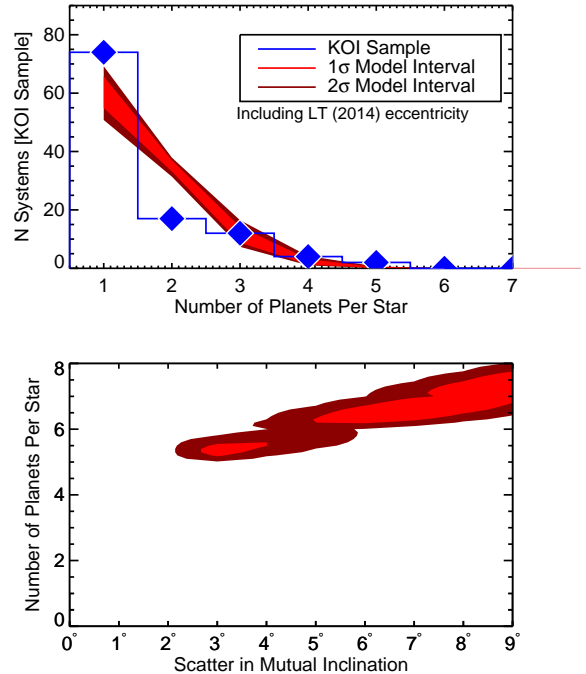} 
 \caption{{\it Top panel}: comparison of the \kepler\ multi--planet yield, in blue, to the best--fitting models at confidences of 1$\sigma$ and 2$\sigma$, depicted by the dark and light red regions respectively. In comparison with Figure \ref{fig:eval_pdf_all}, these models assume that average orbital eccentricity varies with multiplicity as described in \cite{Limbach14}.  {\it Bottom panel}: A contour map of the posterior distribution $P(\{ M\}|N,\sigma)$, with dark and light red corresponding again to 1$\sigma$ and 2$\sigma$ likelihoods.}
  \label{fig:eval_pdf_ecc}
\end{center}
\end{figure}

We then record which planets transit for each iteration to establish a new grid of $\mu(N,\sigma)$, where the eccentricity of planets $e \equiv e(N)$. Our grid is 10 times coarser for this analysis than for our initial test with $e=0$ in order to decrease computational time. We evaluate $\mu$ over a range in $N$ from 1--8 (in increments of 1 planet), and a range in $\sigma$ from 0--9 degrees, in increments of one degree. As expected, the sample of transiting planets is slightly biased toward higher eccentricities than the true underlying distribution, because the closer approach to the star at perihelion increases the transit probability. While this bias is clear for systems with one or two stars, for which eccentricities are on average larger than $e=$0.2, it is negligible for systems with higher multiplicity because the average eccentricity is too low to furnish a much higher likelihood of transit at perihelion. For a single mode of planet formation, we find very similar results as compared to those with fixed eccentricity at zero. In Figure \ref{fig:eval_pdf_ecc}, similar to Figures \ref{fig:eval_pdf_all} and \ref{fig:eval_pdf_multionly}, we show the best-- fitting models to the \kepler\ yield, and contour for the posterior distribution on number of planets and their mutual inclination. We conclude our results are robust from zero to modest eccentricities.

\subsection{Employing a mixture model for a dual population}

We next compare the KOI sample to a two-mode model for planetary architectures. \cite{Hansen13}, in their comparison of the \kepler\ multiple planet yield to their population synthesis models, considered a similar question. They invoked an unknown process that uniformly produced singly transiting systems to compensate for their overabundance. In this case, we define a $\mu^\prime(N,\sigma)$ that is the normalized linear combination of two modes of planet formation: one $\mu_{1}$ defined as having only one planet ($N=1$) that occurs some fraction $f$ of the time, and the other $\mu_{2}$ defined by  $\{N > 1, \sigma_{2}\}$ that occurs with probability $(1-f)$. We adopt the simplified assumption of \cite{Hansen13}, which is that the first mode of planet formation, occurring with probability $f$, produces only singly transiting planets. It's not possible to know, from counting statistics alone, whether these singly transiting systems occurs because less planets exist around the star, or because they are very highly mutually inclined with respect to one another. Indeed, this ``single--planet'' mode is consistent with a system of $N$ planets in which one has a large inclination with respect to the other $N-1$ planets that may have small mutual inclinations. 

\begin{figure}[h!]
\begin{center}
 \includegraphics[width=2.9in]{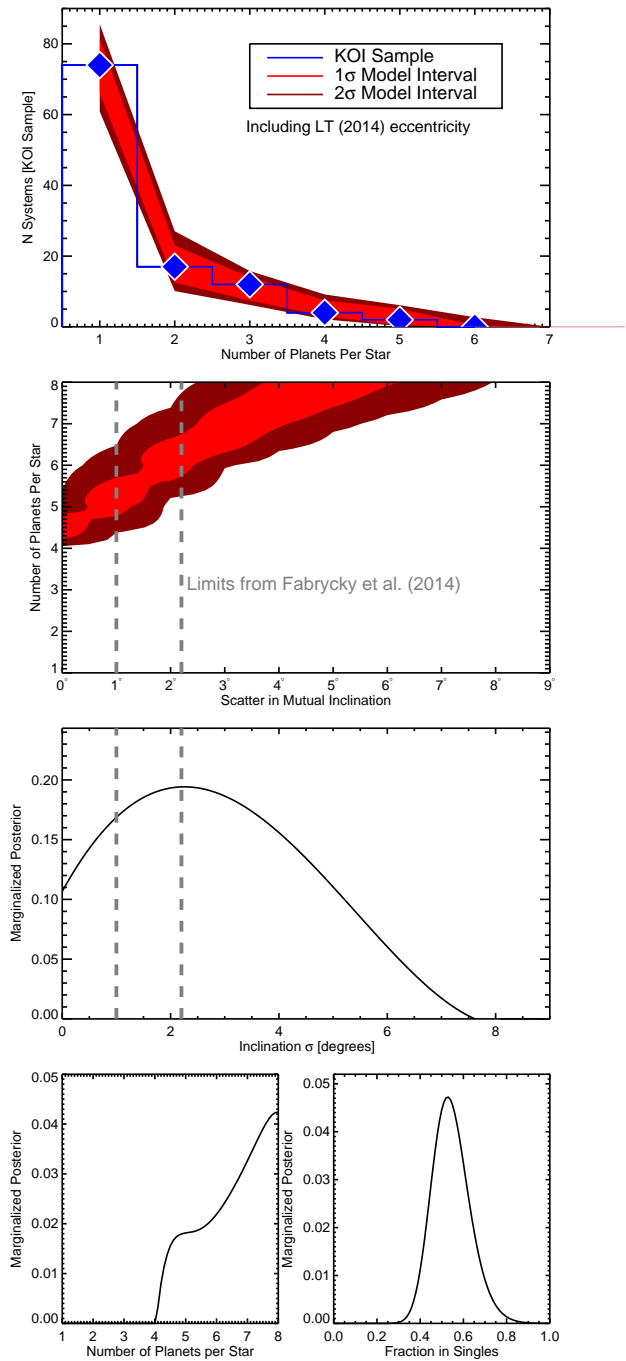} 
 \caption{{\it Top panel}: comparison of the \kepler\ multi--planet yield, in blue, to the best--fitting models at confidences of 1$\sigma$ and 2$\sigma$, depicted by the dark and light red regions respectively. The synthetic planetary population we generated for comparison includes non-zero eccentricities, and allows for a fraction $f$ of the systems to contain a single planet, and a fraction $(1-f)$ to have $N$ planets with mutual inclinations distributed by $\sigma$. {\it Second panel}: A contour map of the posterior distribution $P(\{ M\}|N,\sigma)$, with dark and light red corresponding again to 1$\sigma$ and 2$\sigma$ likelihoods. Gray dotted lines bracket the best-fit range in mutual inclination calculated by \cite{Fabrycky14}. {\it Bottom three panels}: Marginalized posterior distributions (integrated over the other two parameters) for mutual inclination scatter, number of planets star$^{-1}$, and fraction $f$ in excess single systems.}
  \label{fig:eval_pdf_posteriors}
\end{center}
\end{figure}

In this mixture scenario, we can model $\mu_{1}$ as a delta function at $N=1$ transiting planets. We rewrite Equation \ref{eq:bayes1} as follows, now with three free parameters:

\begin{equation}
\mathcal{L} \equiv P(\{M\}|N,\sigma,f)\propto\prod_{n}\frac{\mu^\prime(N,\sigma,f)^{M_{n}}e^{\mu^\prime(N,\sigma,f)}}{M_{n}!},
\label{eq:bayes2}
\end{equation}

where

\begin{equation}
\mu^\prime(N,\sigma,f)=f\cdot\delta(N=1)+(1-f)\mu(N,\sigma).
\label{eq:bayes3}
\end{equation}

\noindent Our results are depicted in Figure \ref{fig:eval_pdf_posteriors}, where we show both the model comparison to the \kepler\ data and the posterior distributions on $N$, $\sigma$, and $f$. In order to reproduce the observed distribution in multiplicity, this unknown mechanism is operating in a fraction $f=0.55^{+0.23}_{-0.12}$ of systems. This is consistent with the estimate of \cite{Hansen13}. They found that if the mechanism generated excess singly-transiting systems operates half the time, with the scenario described in their simulations operating the other half the time, they recover the ratio of doubly-transiting-to-singly-transiting systems of 0.2. We note that our findings are also consistent with \cite{Morton14b}, who considered the ensemble of planets orbiting the \kepler\ M dwarfs. They assumed a single model for M dwarf planetary systems and concluded that each system has 2.0 $\pm$ 0.45 planets star$^{-1}$. This result is in moderate agreement with our value for the {\it average} number of planets per host star, which is 2.8 (assuming 5-planet systems occur 45\% of the time, and 1-planet systems occur 55\% of the time). 

Computing Bayesian evidences in order to test the preference of the data for different models is a complex and subtle undertaking. In this case, the simplicity of our hypothesis enables us to employ a simpler likelihood ratio test, so that we can bypass the evidence calculation. This test is appropriate for our study because we compare the data to a nested family of models, and are testing one branch of the family against another. In our analyses, models with hypothesis (1), in which we have only one mode of planet occurrence, are only a specialized case of models from hypothesis (2), in which a linear combination of two modes contribute. The first set of models can be expressed by setting one of the coefficients in model (2) to zero. For cases with ``nested model likelihoods," such as this one, we are testing the likelihood of one model against a ``special case" of the same model. The test statistic $D$, which encodes how well the data prefers one model over another, is given by 

\begin{equation}
D=-2 \ln\left({\frac{\mbox{Likelihood of model 2}}{\mbox{Likelihood of model 1}}} \right),
\label{eq:ratio}
\end{equation}

\noindent where model 2 is a special case of model 1. We determine that the data prefer the two-mode model by a factor of 21:1. We list our findings in Table \ref{tab:freqs}. 

\begin{deluxetable*}{llllll}
\tabletypesize{\scriptsize} \tablecaption{Best-fit Parameters to \kepler\ M-dwarf Multi-Planet Yield} 
\tablewidth{0pt}
\tablehead{
 \colhead{Data Set}& \colhead{Model} & \colhead{$N$} & \colhead{$\sigma$} & \colhead{$f$} & \colhead{D\tablenotemark{a}}\\
\colhead{ }& \colhead{ } & \colhead{[planets]}& \colhead{[degrees]} & \colhead{ } & \colhead{ } \\}
 \startdata
All data & One mode, $e$=0 & $6.0^{+1.7}_{-1.4}$  & $7.4^{+2.5}_{-5.3}$  & -- & --\\
% & $e$=0 & & & \\
Only multis & One mode, $e$=0 & $6.4^{+1.6}_{-1.6}$ & $4.0^{+2.3}_{-2.9}$ & -- & -- \\
% & $e$=0 & & & \\
All data & One mode, $e\ne0$ & 7.3$^{+0.4}_{-2.1}$ & $8.0^{+1.0}_{-5.0}$ & -- & -- \\
% & $e\ne0$ & & & \\
Only multis & One mode, $e\ne0$ & $>6.9$ & $4.0\pm1.0$ & -- & -- \\
% & $e\ne0$ & & & \\
 &  & & & &  \\
 \hline
  &  & & & & \\
All data & Two modes, $e=0$ & 4.7$^{+3.3}_{-0.6}$ & 1.4$^{+6.3}_{-1.2}$ & 0.58$\pm$0.16 & 14.8 \\ 
% & $e=0$ & & & \\
All data & Two modes, $e\ne0$ & 6.1$^{+1.9}_{-1.9}$ & 2.0$^{+4.0}_{-2.0}$& 0.55$^{+0.23}_{-0.12}$ & 20.8 \\
% & $e\ne0$ & & & \\
\enddata
\label{tab:freqs}
\tablenotetext{a}{See Equation \ref{eq:ratio}. The likelihood of a model with two modes, as compared to a model with one mode.}
\end{deluxetable*}

\section{Investigation of Stellar Hosts}
\label{sec:hosts}

\begin{figure}
 \includegraphics[width=3in]{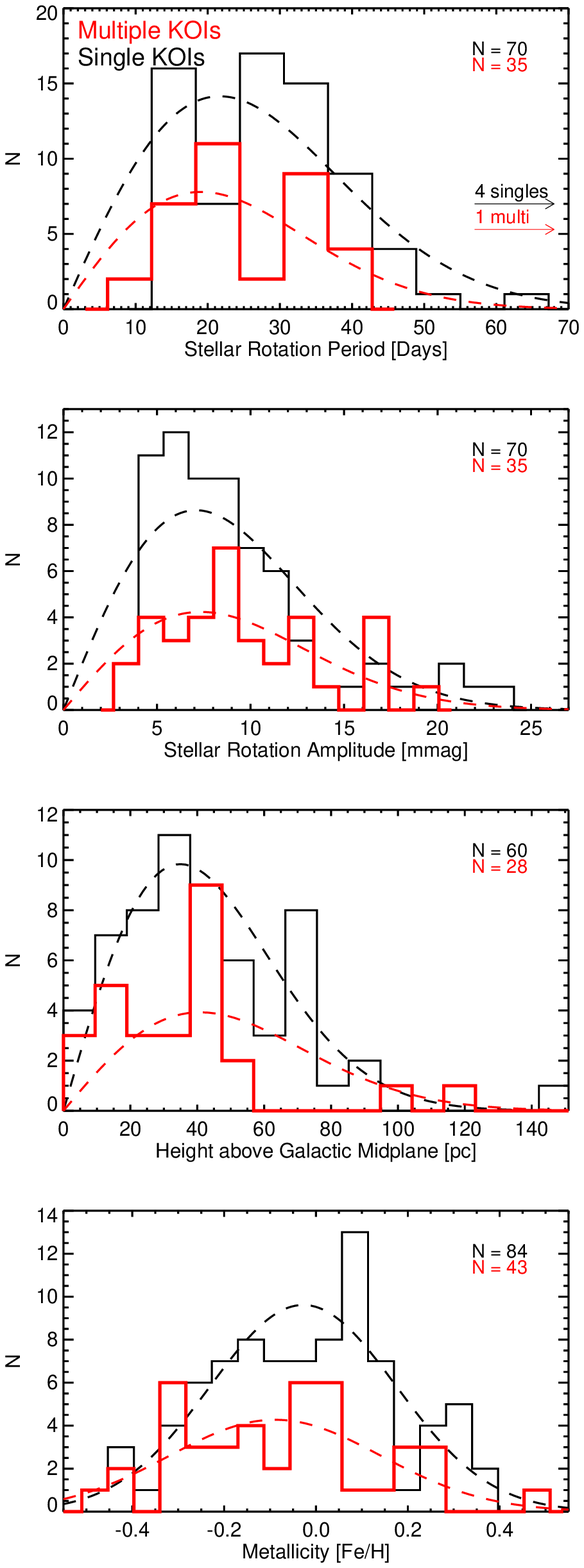} 
 \caption{Comparison of properties for stars hosting one KOI (black), to those hosting multiple KOIs (red). In order from top to bottom panels: stellar rotation period, rotation amplitude, height above galactic midplane, and metallicity. We employ the rotation periods and amplitudes reported by \cite{Mcquillan13a,Mcquillan13b}, the distances from \cite{Muirhead14} to calculate height above galactic midplane, and \cite{Muirhead14} and \cite{Mann14} for metallicities. The dotted curves show the best-fit probability distributions for the single(black) and multiple (red) populations.}
\label{fig:active_multiple}
\end{figure}

We consider whether the host star properties are related to the likelihood of coplanarity or multiplicity. We compare the two planet populations---multi and singleton---in terms of host star rotation period, rotation amplitude, height above the galactic midplane, and metallicity. We use the rotation periods and amplitudes calculated by \cite{Mcquillan13a,Mcquillan13b}. Where available, we use the more recent value (that is, from \citealt{Mcquillan13b}). To calculate galactic height, we employ the distances tabulated by \cite{Muirhead14}. We use metallicities from both \cite{Muirhead14} and \cite{Mann13}. Not all KOIs characterized by \cite{Muirhead14} or \cite{Mann13} also possess a stellar rotation measurement. We indicate the size of each sample in Figure \ref{fig:active_multiple}. There are 5 stars with no detected periodicity in \cite{Mcquillan13a}: 1 of these is a multiple KOI host and 4 others are single KOI hosts. We indicate these stars with lower limit arrows in Figure \ref{fig:active_multiple}. 

Since rotation period, rotation amplitude, and height above midplane are all positively-valued random variables, we model each of these data sets as a Rayleigh distribution, characterized by the scale parameter $\sigma$, e.g. $\sigma_{\rm rot}$ for the spread in rotation periods. We model metallicity, which can take both positive and negative values, as a Gaussian distribution (characterized by a mean $\mu$ and a standard deviation $\sigma$). We assign a Jeffreys prior for both distributions in $\sigma$. \cite{Dey11} show that $p(\sigma)\propto 1/\sigma$ for Rayleigh distributions, identical to the proportionality for the Jeffreys prior in $\sigma$ for Gaussian distributions (in contrast, the Jeffreys prior in $\mu$ is uniform). For the rotation period of the KOIs with one transiting planet, $\sigma_{\rm rot}=21.7^{+1.3}_{-1.2}$ days, whereas for the multiples, $\sigma_{\rm rot}=19\pm1.4$ days. The photometric amplitudes of the rotationally--induced spot modulations are indistinguishable: $\sigma=7.1^{+4.1}_{-3.8}$ mmag for the singles and $7.3^{+0.7}_{-0.5}$ mmag for the multiples. The Rayleigh distribution in galactic height for singly-transiting KOIs is characterized by $\sigma_{\rm H}=35\pm2$ pc and 41$^{+4}_{-3}$ pc for multiples. Finally, the Gaussian distribution in metallicity for the single KOI stars is best modeled by a mean $-0.03\pm0.02$ and standard deviation $0.20\pm0.02$, and for the multiples $-0.09\pm0.04$ and a standard deviation $0.23^{+0.03}_{-0.02}$

\begin{figure}
 \includegraphics[width=3in]{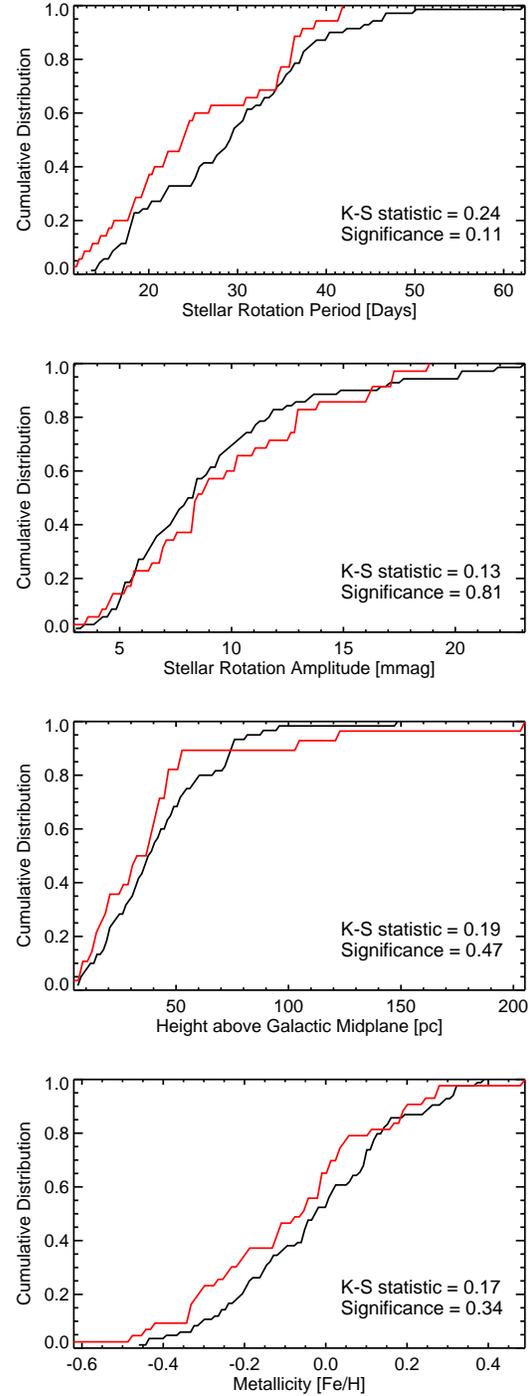} 
 \caption{Cumulative distribution in four host star properties for stars hosting one KOI (black), and those hosting multiple KOIs (red). From top to bottom: stellar rotation period, rotation amplitude, height above galactic midplane, and metallicity. We overplot the K-S test statistic and the significance limit on each panel.}
  \label{fig:mdwarf_tests2}
\end{figure}

\begin{figure}
 \includegraphics[width=3in]{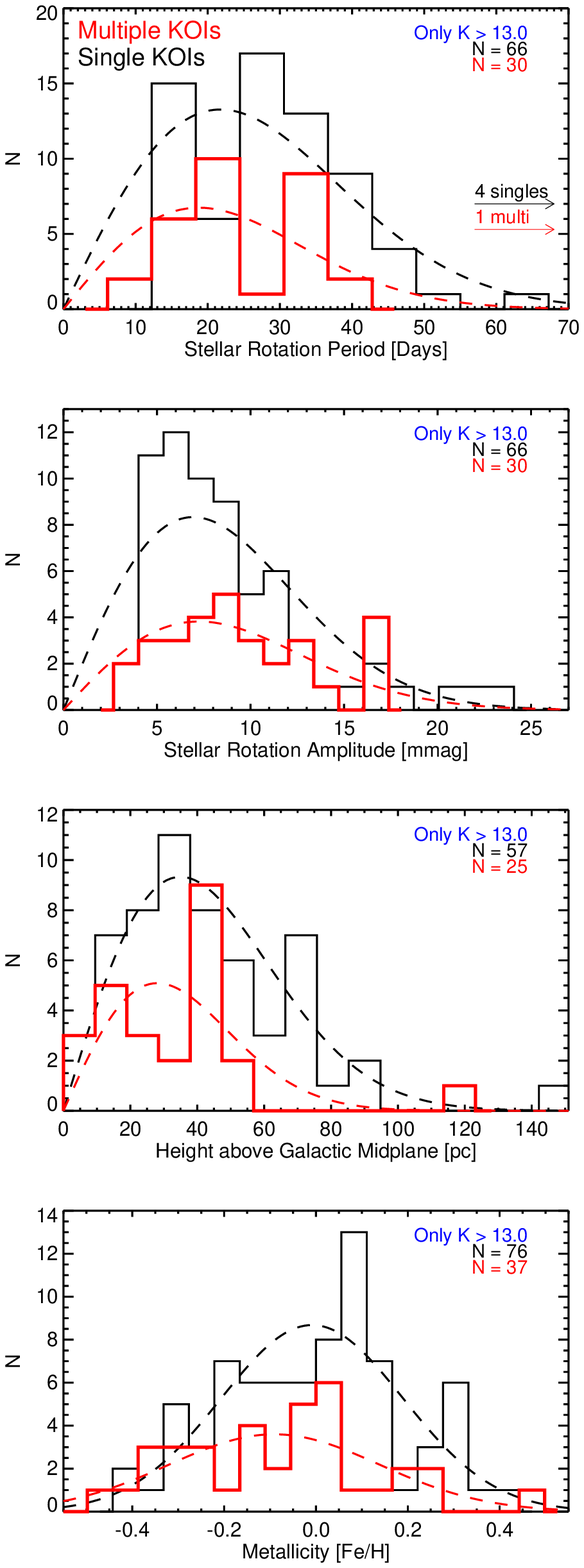} 
 \caption{Identical to Figure \ref{fig:active_multiple}, but for the subset of stars brighter than $K$ magnitude of 13.0.}
  \label{fig:active_multiple_bright}
\end{figure}

\begin{figure}
 \includegraphics[width=3in]{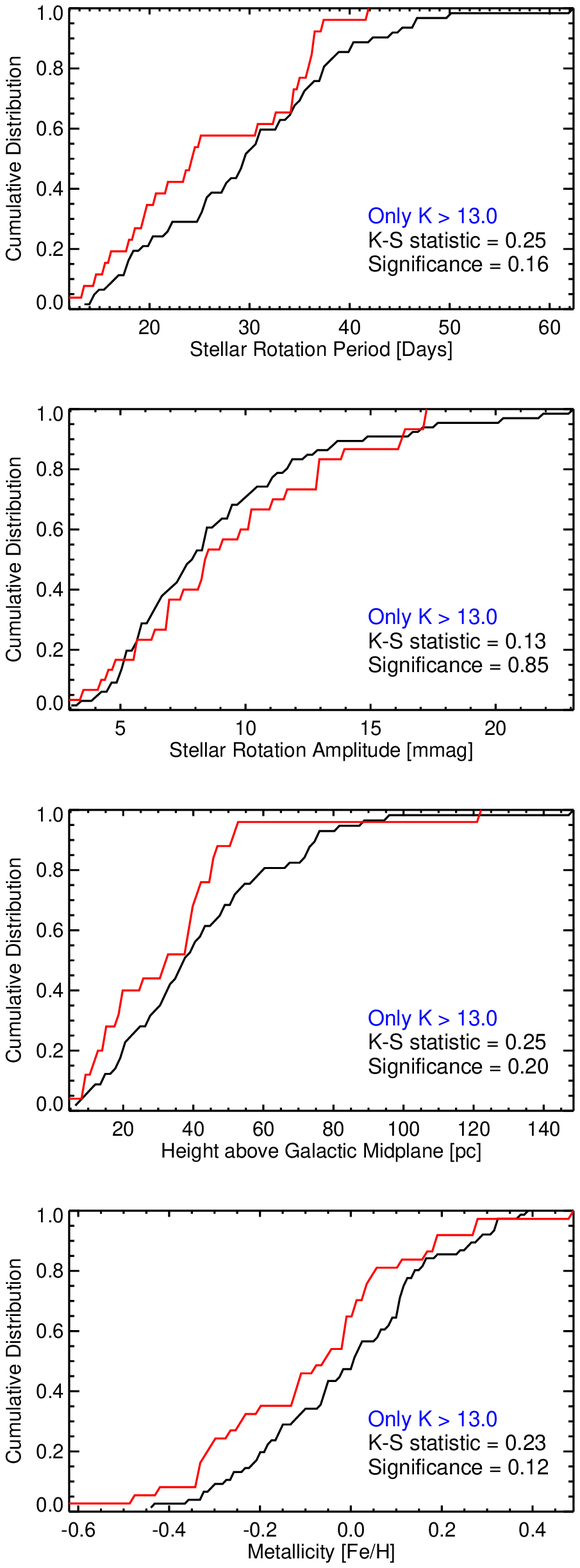} 
 \caption{Identical to Figure \ref{fig:mdwarf_tests2}, but for the subset of stars brighter than $K$ magnitude of 13.0.}
  \label{fig:mdwarf_tests2_bright}
\end{figure}

We quantify the likelihood of each of these properties being drawn from different underlying distributions. We show the cumulative distributions for the singly-transiting and multi-transiting systems in Figure \ref{fig:mdwarf_tests2}. The only parameter for which the K-S statistic exceeds the significance limit is in rotation period (0.24 versus 0.11): the multiply-transiting systems orbit stars that are rotating slightly faster, with marginal confidence (2$\sigma$). 
We also consider the subset of the brightest host stars: those with $K$ magnitudes brighter than 13.0. Here we see trends emerge, with modest significance in three of the four parameters. The K-S statistic exceeds the significance limit in stellar rotation period (0.25 versus 0.16, with hosts of multiple transiting planets rotating more quickly), in height above the galactic midplane (0.25 versus 0.20, with hosts of multiple transiting planets closer to the midplane), and in metallicity (0.23 versus 0.12, with hosts of multiple transiting planets being metal poor). Similarly to Figures \ref{fig:active_multiple} and \ref{fig:mdwarf_tests2}, Figures \ref{fig:active_multiple_bright} and \ref{fig:mdwarf_tests2_bright} show histograms (with best-fit functions overplotted) and cumulative distributions for the four stellar properties. Fitting Rayleigh distributions identically as described above for the subsample of brightest stars, we find $\sigma_{\rm rot}=21.8^{+1.3}_{-1.2}$ days for the rotation periods of the singly transiting planets and $18.8\pm1.4$ days for the multis. The stars with one transiting planet have a typical galactic height of $\sigma_{\rm H}=35.0\pm2.20$ pc and those with multiple transiting planets have a typical height of 28.1$^{+3.0}_{-1.5}$ pc. The multiply-transiting systems are slightly metal poor, with a mean metallicity of [Fe/H]=$-0.09\pm0.04$, in comparison to the singly-transiting systems with average [Fe/H] of $0.0\pm0.02$. The distributions are distinct by 2$\sigma$ in all three cases. We summarize these findings in Table \ref{tbl:stars}.

\begin{deluxetable*}{ccccc}
\tabletypesize{\scriptsize} \tablecaption{Comparison of Stellar Hosts to One or Multiple Transiting Planets} 
\tablewidth{0pt}
\tablehead{
 \colhead{Sample}& \colhead{Stellar Rotation Period} & \colhead{Rotation Amplitude} & \colhead{Height above} & \colhead{[Fe/H]} \\
\colhead{ }& \colhead{[Days]} & \colhead{[mmag]} & \colhead{Galactic Midplane [pc]} & \colhead{ }}
 \startdata
All singles & $21.7^{+1.3}_{-1.2}$ & $7.1^{+4.1}_{-3.8}$ & $35\pm2$ & $-0.03\pm0.02$\\
% & $e$=0 & & & \\
All multis & $19.0\pm1.4$ & $7.3^{+0.7}_{-0.5}$ & 41$^{+4}_{-3}$ & $-0.09\pm0.04$\\
% & $e$=0 & & & \\
 &  & & & \\
 \hline
  &  & & & \\
Brightest singles & $21.8^{+1.3}_{-1.2}$ & 6.9$\pm$0.4 & $35.0\pm2.2$ & $0.0\pm0.02$\\
% & $e\ne0$ & & & \\
Brightest multis & 18.8$\pm$1.4 & 7.2$^{+0.4}_{-0.6}$ & 28.1$^{+3.0}_{-1.5}$ & $-0.09\pm0.04$ \\
% & $e\ne0$ & & & \\
\enddata
\label{tbl:stars}
\end{deluxetable*}

\section{Conclusions}
\label{sec:conclusions}

We have investigated the \kepler\ multiplicity yield of planets transiting M dwarfs. By comparing to predicted yields, given M dwarfs with $N$ planets star$^{-1}$ and an average mutual inclination $\sigma$, we make the following conclusions:

\begin{itemize}
\item The multiplicity statistics of the \kepler\ exoplanet sample orbiting M dwarfs cannot be reproduced by assuming a single planetary system architecture. The best-fitting models, which have large multiplicities ($N>5$) and large mutual inclinations ($7^{\circ}$), don't resemble the typical \kepler\ multi-planet systems, which are mutually inclined by $1-2.2^{\circ}$ \citep{Fabrycky14}. 

\item However, the multi-planet yield (that is, the sample of KOIs hosting 2 or more transiting planets) is well-modeled by a well-populated and approximately coplanar architecture. The best-fitting planetary model has 6 planets, and typical mutual inclination of 2$^{\circ}$. Because M dwarf stars are the most populous in the Galaxy, we consider this architecture a ``typical" galactic multi-planet system \citep[cf also][]{Swift13}. 

\item These conclusions are robust to assumptions of eccentricity. We report approximately the same findings whether we force $e=0$ for all planets, or whether we drawn eccentricities from the distributions of \cite{Limbach14}.  

\item If we invoke an additional population of singly-transiting planets, this dual population is preferred by a factor of 21 to that with only one type of planetary architecture. These excess singly transiting planet systems are 0.55$^{+0.23}_{-0.12}$ of the total number of systems; half of systems are in well-aligned multies, and half are either singleton or in systems with large mutual inclinations.

\item Three stellar properties, with modest significance (95\% confidence), are predictive of whether a given M dwarf hosts a single planet, or multiple planets. These are stellar rotation, height from galactic midplane, and metallicity: multiply-transiting systems are on average rotating more quickly, closer to the midplane, and metal poor. This finding holds when the investigation includes the brightest host stars ($K>13.0$), which is 80\% of the total sample.  
\end{itemize}

Though individually the properties of the stellar hosts to multiple transiting planets are only modestly different from those hosting a single transiting planet, the potential implications are intriguing. The deficiency in metals in among the multiples would be consistent with the framework set forth by \cite{Dawson13}. In that study, they found that Jovian planets orbiting metal-rich stars have more eccentric orbits on average. They posit that metal-rich circumstellar disks give rise to multiple giant planets, which then could scatter to larger eccentricities. In this sense,  metal-rich stars host dynamically ``hotter" planets on average, and so we ought to observe flatter systems around metal-poor stars, as we tentatively report. This mechanism would also be consistent with the fact that the largest planets orbiting M dwarfs in the \kepler\ sample ($>$ 6 $R_{\oplus}$) are all singles, as we note in Section \ref{sec:bias}. By two common metrics of youth in M dwarfs (namely, higher rotation rate and proximity to galactic midplane, per e.g. \citealt{West06}), the distribution of stars hosting multiple transits is younger on average. It is possible that stellar age is predictive of planetary architecture for the stars in our sample. If so, the mechanism that operates to disrupt the coplanar set of planets must occur after the short-lived ($\sim10$~Myr) ``planet formation" epoch during which there exists a massive gas disk. 

One as-yet untested parameter, which may be predictive of planetary architecture, is stellar binarity. The disruption of the circumstellar disk by a nearby companion may result in the departure from the compact and coplanar distribution observed in half the sample. Alternatively, such systems may form fewer planets. Dynamical disruption of some kind seems the likeliest scenario, since \cite{Morton14} demonstrated that the orbits of the singly-transiting systems are more misaligned with the spin of the host star. If binarity were responsible in full for the excess population of singly-transiting planets, then we might simplistically require 50\% of M dwarfs to reside in binaries. This is a reasonable estimate for solar-type stars, for which 50\% exist in binary systems \citep{Duquennoy91}.

This reasoning assumes that each M dwarf (whether in a single or multiple star system) is as likely to host planets as another. \cite{Fischer92} measured a multiplicity rate of $42\pm9$\% among M dwarfs over a wide range of orbital separations, while \cite{Leinert97} and \cite{Reid97} found that the binary frequency for M dwarfs is only 30\%. Recent work by \cite{Janson12} concluded that only 27$\pm$3\% of M dwarfs are part of binary systems with separations between 3--227 AU. 

However, even the assumption of constant planet formation independent of binarity is not well-supported: \cite{Wang14} found that binarity can decrease planetary formation by a factor of several, with modest confidence. In that work, the impact on planet formation decreases with increasing orbital separation of the stars (with 10 AU being the closest orbital distance in their investigation). Intuitively, closer binaries are likelier to produce disrupted disks, and only 3--4\% of M dwarfs have binary companions closer than 0.4 AU (an orbital period of 230 days for a planet orbiting an M0 dwarf) per \cite{Clark12}.  It's therefore not plausible to invoke binarity as the complete explanation for excess singly-transiting systems, though it may play some role. 

We require more exoplanetary systems  to robustly test whether any of the suggestive properties we have identified in this work inform M dwarf planetary systems. In the next upcoming years, such a sample will be forthcoming from NASA's Two-Wheeled extended \kepler\ Mission, known as K2 \citep{Howell14}. Accepted GO programs\footnote[2]{http://keplerscience.arc.nasa.gov/K2/GuestInvestigations.shtml, accessed on 4 October 2014} include more than 15,000 M dwarfs in Campaigns 1--5, with each Campaign field receiving 75 days of continuous monitoring. In particular, the trend with Galactic height will be better tested by K2 than by the nominal mission, because fields in the first Campaigns include both the galactic plane and the poles. The addition of a fresh sample of M dwarf planets will allow firmer conclusions about which stellar properties sculpt their planetary architectures. 
\newpage
\newpage

\bibliographystyle{apj} 
\bibliography{allrefs}

\newpage

\end{document}